\documentclass[11pt,a4paper]{emulateapj}
\usepackage{epsfig}
\usepackage{amsmath}
\usepackage{natbib}

 \newcommand{\ostrv}{\O12}

\newcommand{\rvlin}{{\sc rvlin}}
\newcommand{\iraf}{{\sc iraf}}
\newcommand{\idl}{{\sc idl}}

\begin{document}

\title{A Radial Velocity Study of Composite--Spectra \\ Hot Subdwarf 
  Stars with the Hobby--Eberly Telescope\footnotemark[*]}
\footnotetext[*]{Based on observations obtained with the Hobby--Eberly
  Telescope, which is a joint project of the University of Texas at
  Austin, the Pennsylvania State University, Stanford University,
  Ludwig--Maximilians--Universit\"{a}t M\"{u}nchen, and
  Georg--August--Universit\"{a}t G\"{o}ttingen.}

\author{Brad N. Barlow$^1$\footnotemark[\dag], Richard A. Wade$^1$, Sandra E. Liss$^1$, Roy H. \O stensen$^2$, \& Hans Van Winckel$^2$}
\footnotetext[\dag]{bbarlow@psu.edu}

\affiliation{$^1$Department of Astronomy \& Astrophysics, The Pennsylvania State University, 525 Davey Lab, University Park,
PA 16802\\ 
$^2$Instituut voor Sterrenkunde, K.U. Leuven, B-3001 Leuven, Belgium}


\submitted{Received 2012 May 31; accepted 2012 August 17; published 2012 September 26} 

\begin{abstract}
  Many hot subdwarf stars show composite spectral energy distributions
  indicative of cool main sequence companions. Binary population
  synthesis (BPS) models demonstrate such systems can be formed via
  Roche lobe overflow or common envelope evolution but disagree on
  whether the resulting orbital periods will be long (years) or short
  (days). Few studies have been carried out to assess the orbital
  parameters of these spectroscopic composite binaries; current observations suggest the
  periods are long. To help address this problem, we selected fifteen
  moderately-bright (V$\sim$13) hot subdwarfs with F--K dwarf companions
  and monitored their radial velocities (RVs) from January 2005 to July
  2008 using the bench--mounted Medium Resolution Spectrograph on the
  Hobby--Eberly Telescope (HET). 
  Here we describe the details of our observing, reduction, and analysis
  techniques and present preliminary results for all targets.  By
  combining the HET data with recent observations from the Mercator
  telescope, we are able to calculate precise orbital solutions for
  three systems using more than 6 years of observations.  We also
  present an up--to--date period histogram for all known hot subdwarf
  binaries, which suggests those with F--K main sequence companions tend
  to have orbital periods on the order of several years.  Such long
  periods challenge the predictions of conventional BPS models, although
  a larger sample is needed for a thorough assessment of the models'
  predictive success. Lastly, one of our targets has an eccentric orbit,
  implying some composite--spectrum systems might have formerly been
  hierarchical triple systems, in which the inner binary merged to
  create the hot subdwarf.
\end{abstract}

\keywords{ binaries: spectroscopic --- ephemerides --- subdwarfs --- techniques: radial velocities}

\section{Introduction}
\label{sec:intro}

Binarity plays an essential role in the story of the hot subdwarf B
(sdB) stars, one of the most enigmatic stages of stellar evolution.  The
sdBs dominate surveys of faint blue objects and are found in almost all
Galactic stellar populations. They are the field counterparts of the
Extreme Horizontal Branch stars in globular clusters, and
their location in the H--R diagram corresponds to stellar models with
He--burning cores and extremely thin H envelopes \citep{heb86}.
Presumably, the sdB progenitors were stripped of almost all their
surface hydrogen while on or near the red giant branch (RGB), 
igniting helium burning in their cores at nearly the same time.
The question is how this can happen, since most Population I stars do not
lose their entire envelopes, and remain in the red clump, not far from
the RGB, during core He burning.

 \begin{table*}
 \centering
  \caption{RV Monitoring Targets} 
\scriptsize
  \begin{center}
    \leavevmode
    \begin{tabular}{lllllll} \hline \hline              
Target & Sp. Type$^a$ &RA & Dec & V & Alternate &Comments\\
             &                  & [J2000] & [J2000] & & Name & \\
\hline
PG 0039+049 &  sdB+G2V 	&  00:42:06.1	&+05:09:24 	&   12.9	&  PB 6107 &  
\\
PG 0110+262 & sdB+G0V   		& 01:13:14.9   	&+26:27:31  	&   12.9   	&    & \\
PB 8783 & sdOV$^{b}$+F4V  		& 01:23:43.2  	&  -05:05:45   	&    12.3    & EO
Cet &  pulsating hot subdwarf$^5$ \\  
PHL 1079 & sdB+G7V		& 01:38:27.1  	& +03:39:39   	& 13.4  	&   &          \\  
PG 0232+095 & sdB+G1V 		& 02:35:12.0   	&+09:45:38	&  12.5    	&      &\\  
PG 1040+234 & sdB+G1V 	&10:43:39.3  	& +23:09:07   	& 13.4   	& TON 1281&  
resolved$^{1}$    \\  
PG 1104+243 &  sdOB+G2V 	& 11:07:26.3   	&+24:03:12   	& 11.3  	&  &  MS orbit
published$^4$\\   
PG 1206+165 & sdB+G9V$^c$	& 12:09:16.7   	&+16:11:56   	& 13.8    	&  PB 3854 &
\\  
PG 1253+284 &sdB+F8V	& 12:56:04.9   	&+28:07:20 	& 12.7   	& TON 139    &sdB
orbit published$^{3}$; resolved$^{1}$; presumed triple system$^{1,3}$ \\ 
PG 1317+123 &  sdO+G1V	& 13:19:53.6   	&+12:03:59   	& 11.3     	 & Feige 80  &
MS orbit published$^4$  \\ 
PG 1338+611 &  sdB+G4V 	&13:40:14.7   	&+60:52:48      	&11.4	&   Feige 87   &
MS orbit published$^4$    \\
PG 1449+653 & sdB+G5  	&14:50:36.1   	&+65:05:53   	&13.6	&   &    \\  
PG 1629+081 &  sdOB+K5V$^c$ 	& 16:32:01.4   	& +07:59:40   	& 12.8	&   &
resolved$^{2}$  \\   
PG 1701+359 & sdB+G8	& 17:03:21.6 	&  +35:48:49   	& 13.2	&   &    \\  
PG 1718+519 & sdB+G4V	&17:19:45.5  	& +51:52:10 	& 13.7	&     &  resolved$^{1}$
     \\   
\hline
\multicolumn{7}{l}{$^{a}$classifications from this work (see \S
\ref{subsec:classifications}) unless otherwise noted}\\
\multicolumn{7}{l}{$^{b}$sdO status of the hot subdwarf recently confirmed by
\citet{ost12b}}\\
\multicolumn{7}{l}{$^{c}$companion classification from \cite{staphd}}\\
 \multicolumn{7}{l}{\textbf{References}: $^{1}$\citet{heb02};
$^{2}$\citet{ost05}; $^{3}$\citet{cop11}; $^4$\citet{ost12}; $^5$\citet{koe97} 
      }\\ 
    \end{tabular}
  \end{center}
  \label{tab:targets}
\end{table*}

Stochastic mass loss from single stars \citep{dcr96} or other
single--star scenarios to explain this are \textit{ad hoc}, so most
recent effort has concentrated on mass loss resulting from interactions
in a close binary star system, following the early exploration by
\citet{men76}. \citet{han02,han03}, for example, described five channels
of binary star evolution that can lead to the formation of hot
subdwarfs: the `first' and `second' stable Roche--lobe overflow (RLOF)
scenarios, the `first' and `second' common envelope (CE) channels, and
the merger of two He-core white dwarfs (WDs).  Their CE scenarios produce
close binaries with orbital periods from hours to days; the sdB
companions are generally late-type (G/K/M) main sequence stars or white
dwarfs.  In contrast, their stable RLOF channels lead to systems with
early--type (B/A/F) main sequence or late--type giant companions and
long orbital periods ($P \sim$\, few $\times\ 100$ d).  An alternative
scenario employing the $\gamma$--formalism has been put forth by
\citet{nel00,nel01a} and \citet{nel10}, whose CE channels can produce
systems with main sequence companions and periods on the order of years.

To date, orbital parameters have been measured for approximately 150 sdB binary
systems; the overwhelming majority of these are close binaries with $P <
15$ d (see Table A.1 of \citealt{muchfuss} for a recent summary).  The
companions are mostly WDs and M dwarfs, essentially invisible
in the glare of the sdB.  With a suitable choice of tuning parameters,
theoretical models can reproduce the observed distribution of sdB masses
and orbital periods in these systems reasonably well
\citep{max01,cop11}.  The recent binary population synthesis (BPS) study
by \citet{cla12}, however, shows that numerous parameter sets can
reproduce this observed subpopulation; thus, \textit{additional
  constraints are needed to better tune BPS codes}.

Very little attention has been paid to the sdB+F/G/K
binaries, in spite of the nearly equal flux contributions from the
two components in such systems. The relative dearth of published orbital
parameters implies their periods are quite long.  More than a decade
ago, \citet{gre01} hinted that sdB binaries with composite spectra tend
to have longer orbital periods.  They reported an average around 3-4
years, although no specific systems were named. Since that brief
discussion, only two other studies have claimed the detection of orbital
periods in composite--spectrum binaries, both of which were published very recently.  
\citet{ost12} (hereafter \ostrv) discuss a survey
conducted with the High Efficiency and Resolution Mercator Echelle Spectrograph
(HERMES, see \citealt{ras11}) mounted at the Mercator Telescope on 
La Palma, Spain and present preliminary results showing
orbital periods longer than $\sim$ 500 d for at least eight binaries.
In addition to that work, \citet{dec11} report a period of 760 d for PG
1018$-$047.

Orbits with such long periods are sufficiently large to have once
accommodated an inner binary, so it is possible that these systems were
formerly hierarchical triple systems, in which case the subdwarf could
have formed from the merger of the inner binary.  \citet{cla11}, for
example, recently proposed a scenario in which the merger of a He--core WD 
and a low--mass main sequence star could eventually form an
sdB.  If such a binary were in orbit with a dwarf companion, the merger
would leave behind an sdB and a main sequence star that had no part in the
formation of the sdB (aside from potentially advancing the merger
process via the Kozai mechanism).  Unlike stable RLOF-- and CE--produced
systems, which should have nearly circular orbits, no limitations exist
on the eccentricities of these binaries other than a requirement that
the periastron separation not be too small.  Thus, it is imperative to
constrain the orbital geometry in addition to the period and radial
velocity (RV) amplitudes.


Here we describe a survey carried out with the Hobby--Eberly Telescope
(HET) over $\approx 3.5$ years to measure the line--of--sight accelerations of a sample
of sdBs with composite spectra.  We discuss our target selection in \S
\ref{sec:targets} and observing strategies in \S \ref{sec:obs}.  The
details of our analysis techniques for determining velocities and
fitting the RV curves are given in \S \ref{sec:analysis}.  Preliminary
results and the current status of our program are presented in \S
\ref{sec:results}.  After combining the HET data with recent
measurements from \ostrv , we report precise orbital solutions for three
of the targets and discuss them in greater detail.  Our initial results
confirm early suggestions that the orbital periods of many sdB+F/G/K
systems are long, on the order of several years.  We present in \S
\ref{sec:histogram} an updated orbital period histogram for all measured
sdB binaries, which shows a possible gap in the period distribution
centered near $P=70$ days, and we briefly discuss the puzzle that it
presents and possible resolutions.  Finally, we summarize our results
in \S \ref{sec:conclusions}.


\section{Target Selection}
\label{sec:targets}

Our criteria for selecting targets were simple: targets were required to
be (i) hot subdwarfs with late F, G, or early K dwarf companions, (ii)
relatively bright (V$\approx$11--14), and (iii) observable from the HET,
situated at McDonald Observatory.  F/G/K main sequence companions have
similar enough luminosities to hot subdwarfs that these systems show
composite spectra at optical and/or infrared wavelengths.  The cool
companions' rich spectra provide a plethora of lines for precise RV
measurement and match the solar spectrum (G2\,V) well enough that twilight
sky spectra obtained nightly can be used as high signal--to--noise (S/N) ratio
RV templates.  Informed by previous observations from \citet{gre01}, we
selected fifteen binaries meeting the above criteria, as presented in Table
\ref{tab:targets}.  All targets were recognized as potential hot
subdwarf systems in the Palomar--Green Survey \citep{gre86}, except for
PHL 1079 \citep{har62,kil84} and PB 8783 \citep{ber84}.  Rough spectral
types for the cool companions have been published for most of the
targets, but since these come from different sources and sometimes differ
significantly, we classified them ourselves as described in \S
\ref{subsec:classifications}

Several of the targets have been observed in greater detail already.
\citet{heb02} and \cite{ost05}, for instance, found that four of the
systems are resolved binaries with separations up to 2\arcsec.  Of
these, PG 1253+284 has been recognized as a potential triple--star
system due to the detection of a rapid acceleration of the sdB star that
is incompatible with the separation implied by the 0\arcsec.32 angular
distance between the two observed components.  PB 8783 exhibits
stellar pulsations \citep{koe97} with surface velocities that complicate
the characterization of any orbital accelerations around its F--type
companion.  Most notably, three of our targets (PG 1104+243, PG
1317+123, and PG 1338+061) were recently observed in an RV monitoring
program similar to ours by \ostrv, who report circular orbit solutions
from measurements of the cool companion lines.  In \S \ref{sec:results},
we combine the HET data with their more recent results to gain
a total observing baseline of $\sim$6 yr for these binary systems.

\section{HET Observations and Reductions}
\label{sec:obs}
Over a three--year period from 2005 to 2008, we monitored all survey
targets spectroscopically with the bench--mounted Medium Resolution
Spectrograph (MRS; \citealt{ram98}) on the 9.2-m HET.  The HET
operates in a queue--scheduled mode.  Follow-up observations of some of
our targets are currently being carried out with the High Resolution
Spectrograph (HRS) on HET and will be discussed in a future paper.  We
used the ``blue'' 1.5\arcsec\ optical fiber pair, the 220 mm$^{-1}$
cross-disperser grating, and 2$\times$2 on--chip binning to achieve an
average resolution of $R$=10,000 over the 4400--6200 \AA\ spectral
range.  This configuration yields an average dispersion of $\simeq$ 10
km s$^{-1}$ per binned pixel and samples $\sim$3.2 pixels per resolution
element.  The pair of fibers, which are separated by 10\arcsec, allowed
us to monitor the sky flux and target flux simultaneously.  Exposure
times ranged from 180 to 1800 s depending on stellar magnitude, giving
us a typical S/N ratio of 80 per resolution element.
 
Each object spectrum was preceded and/or followed by a pair of ThAr
comparison lamp spectra (with different exposure times) for wavelength
calibration.  Standard calibration frames were collected each night,
including twilight sky spectra, bias frames, and quartz lamp flat--field
spectra.  On most nights we also observed telluric, RV, and
spectrophotometric standard stars using the same instrumental setup.

We used the \textit{ccdproc} routine in {\sc iraf}\footnote{IRAF is
  distributed by the National Optical Astronomy Observatories, which are
  operated by the Association of Universities for Research in Astronomy,
  Inc., under cooperative agreement with the National Science
  Foundation} to bias--subtract and flat--field the spectra, then
optimally extracted each aperture with the \textit{apall} function.
Finally, we used the \textit{ecidentify} routine 
in {\sc iraf} to identify comparison lamp emission lines and construct wavelength 
solutions. The ThAr solutions typically used $\sim$ 100 lines and
resulted in an RMS scatter of $\sim$ 300 m s$^{-1}$ per line
(approximately one thirtieth of a pixel).

After discarding orders severely contaminated by sky emission lines, CCD
fringing, or cross--disperser order overlap, as well as crowded or
low--throughput orders, we found the useful wavelength coverage to be
$\sim$4400--6200 \AA.  Since the data were taken without dividing the
exposures to aid cosmic ray removal (so--called ``splits''), we
developed an {\sc idl} code to detect and record the locations of cosmic
ray events in the extracted spectra.  We made no attempt, however, to
remove these or adjust the flux in these pixel locations but simply
ignored compromised wavelength bins during the analysis that followed.

\section{Analysis Methods}
\label{sec:analysis}

\subsection{Classifying the cool companions}
\label{subsec:classifications}

We estimated the spectral types of the cool companions by comparing
the observed HET spectra to high--resolution templates in the ELODIE 3.1
stellar spectral library \citep{pru01}.  Spectral types F0--M9 were
considered in luminosity classes IV and V.  Using our own {\sc IDL} 
code, we compared the best individual spectrum for each target to
the template spectra over the wavelength range 5075--5200 \AA, which
includes important discriminators such as the \ion{Mg}{1}{\em b}
triplet (+MgH) and Fe lines.  Before any comparisons were
made, we degraded the resolution of the ELODIE templates (R=42,000) by
convolving them with Gaussians to match the resolution of the HET/MRS
observations (R=10,000).  We constructed a library of possible
``composite--spectrum'' models, diluting each ELODIE template spectrum
by a flat continuum (representing the contribution from the hot
subdwarf) using a range of dilution factors (5--95\%).  We then normalized
 the observed MRS spectra with
high--order polynomials using the \textit{continuum} function in {\sc
  IRAF} and shifted each spectrum to its
rest--frame.

For each object--template pair we subtracted the model from the object
and computed the sum of squared residuals (SSR); the adopted
best--fitting dilution is the one that minimizes the SSR. 
 We repeated this process for all ELODIE templates
and assigned a final spectral type for each cool companion target
corresponding to the global SSR minimum. To test the accuracy of our
fitting technique, we also classified several twilight--sky spectra
and a dozen RV standards with known spectral types (drawn from the
{\it Astronomical Almanac} and \citealt{nid02}).  Our
results matched the known values reasonably well, falling within two
subtypes of the published classification 80\% of the time.  The
largest disagreement 
encountered in our consistency check was 4 subtypes.

The second column in Table \ref{tab:targets} lists our classifications
for the target stars, which range from F4\,V to G8\,V.  With 80\%
confidence, they are accurate to within 2 subtypes.  They are also
reasonably consistent with previously published values (e.g.,
\citealt{staphd}).  Although we report all companions as dwarfs, two of
the targets, PG 0039+049 and PG 1317+123, are matched equally well with
subgiant templates.  Classifications could not be made for PG 1206+165
(companion lines too weak) or PG 1629+081 (resolved binary, companion
not on fiber).  In these cases, we adopt the spectral types reported by
\citet{staphd}.

\subsection{Measuring the radial velocities}
\label{subsec:measuring_RVs}
  \begin{figure}
  \begin{center}
   \includegraphics[scale=1]{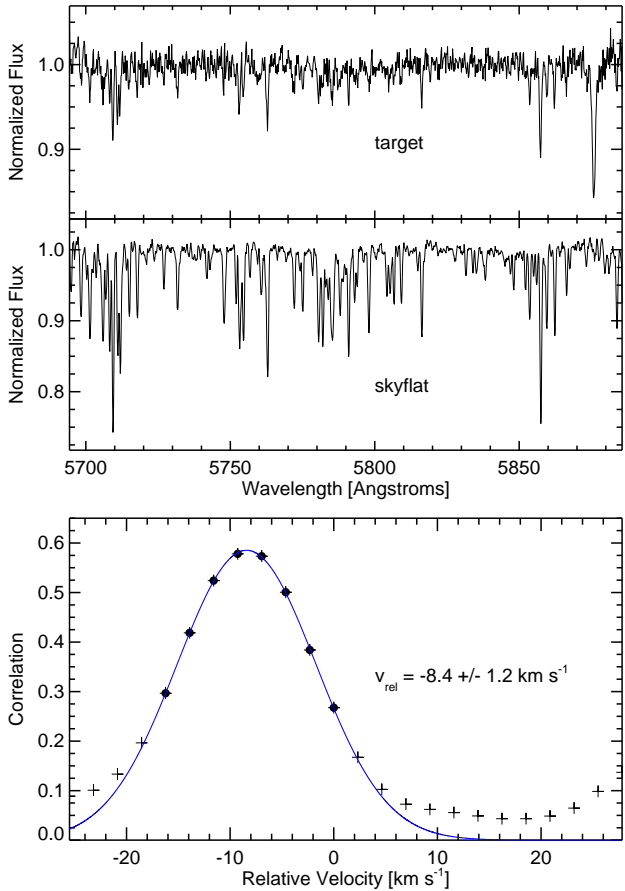}
   \caption{Example cross--correlation of a target--sky spectrum pair.  A
     single order of one of the PG 1104+243 spectra (top panel) is shown
     above the same order of a twilight--sky spectrum (middle panel).
     The \ion{He}{1} 5876 \AA\ line from the sdB star stands out in the target
     spectrum amongst a plethora of features from the cool companion.
     The cross--correlation function (bottom panel) shows a reasonable
     match between the cool companion and solar features.  The blue line
     shows the best--fitting Gaussian to the correlation values marked with
     solid points.}
     \label{fig:cross_correlation}
  \end{center}
\end{figure}

We used \iraf\ routines to find RVs of the targets.  Our primary
strategy was to exploit the abundance of spectral features from the cool
companion by cross--correlating each spectrum against a twilight sky
(i.e., solar) spectrum obtained on the same night.  Before this was done, we
created sampling region files for each spectrum that defined acceptable
wavelength bins to use for the velocity fitting.  In addition to
ignoring bins compromised by cosmic rays and sky emission features, we
also avoided spectral regions heavily influenced by absorption lines
from the hot subdwarf component, namely the Balmer lines (H$\beta$,
H$\gamma$), \ion{He}{1} lines (4472, 4713, 5015, 5876 \AA) and the
\ion{He}{2} 4686 \AA\ line.  Using only these data samples, we used the
\iraf\ \textit{continuum} task to fit the continuum and normalize it to
unity.  We then cross--correlated the normalized spectra against the
solar spectra, order--by--order, using \textit{fxcor} and
\textit{rvcorrect} to determine heliocentric RVs.

Figure \ref{fig:cross_correlation} shows an example target--sky pair
and its cross correlation function.  Most targets matched the solar
spectrum with a correlation nearing 0.8, although systems with
companions deviating significantly from G2\,V had lower values. 
  Our fitting process resulted in $\sim$14 RV
measurements per target observation, depending on the exact number of
echelle orders used.  RV errors were computed from the
standard deviation of these values about the mean.  We find typical errors
 of 600--1100 m s$^{-1}$.  

In some cases, it was also possible to estimate the sdB star's RV from
the available \ion{He}{1} lines and, for the hotter stars, \ion{He}{2}
4686 \AA. We avoided using H$\beta$ for this purpose since its profile
spans almost an entire echelle order; also, its core is blended with
the H$\beta$ profile of the cool companion.  For each of these
targets, after selecting sampling regions around the sdB He lines, we
cross--correlated each spectrum against a self--template, chosen to be
the highest S/N spectrum obtained for the target.  This process only
provides relative RV measurements with respect to this template;
converting them to absolute velocities requires determining the
heliocentric velocity of the template.  Fitting simple Gaussians or
other symmetric functions to the sdB lines cannot be used for this
purpose since the \ion{He}{1} lines are asymmetric and would skew the
measured centroids.  Instead, we cross--correlated the observed
profiles against a model sdB spectrum with $T_{\rm eff}$ = 30,000 K,
$\log g = 5.6$, and $\log {\rm N(He)/N(H)} = -2$.  We then adjusted the
relative measurements
appropriately. Naturally, the RV precision achieved from this process
is not ideal, but it was sufficient to detect and measure sdB
accelerations in many of the systems.  RVs determined from \ion{He}{1}
lines may show a systematic offset, however, as some of the lines are
sensitive to pressure shifts in the high--gravity atmosphere
(especially the 5876 \AA\ D$_3$ line).  Additionally, the
gravitational redshift difference between the cool companion and the
compact hot subdwarf can give the appearance of a $\sim$2 km s$^{-1}$
offset.  These issues have been encountered before in studies of hot
subdwarfs (e.g., \citealt{oro97}) and also DB white dwarfs (e.g.,
\citealt{koe87}), where the offset can approach tens of km
s$^{-1}$.  We will discuss their effects on our results 
object--by--object, as needed.

To investigate the zero--point accuracy of the MRS velocities, we
cross--correlated all of the observed RV standards against
twilight sky spectra (usually taken on the same night) 
and compared our results to the \citet{nid02}
values. A histogram of the RV differences is described reasonably well
by a Gaussian distribution with a dispersion of $\sigma \approx 340$ m
s$^{-1}$; its centroid is within 100 m s$^{-1}$ of zero. We performed a
similar test using all of the available twilight sky spectra,
cross--correlating them against the highest S/N twilight sky spectrum. In
this case, we find a scatter with $\sigma$ = 250 m s$^{-1}$. Our RV
standard star scatter is roughly consistent with the twilight sky result,
once the additional scatter of $\approx 150$ m s$^{-1}$ from the
\citet{nid02} measurements is added in quadrature.

\subsection{Fitting the orbital parameters}

We fitted each RV curve with a model of the form:

\begin{equation}
V_{\rm rad}= \gamma + K [\cos(\theta + \omega) + e \cos(\omega)]
\end{equation}
where $\omega$ is the periastron angle, $K$ the RV semi--amplitude,
$\gamma$ the systemic velocity, and $e$ the eccentricity. The true
anomaly ($\theta$) is defined in terms of the period ($P$),
eccentricity, and time of periastron passage ($T_0$) through the
relation

\begin{equation}
\tan \left(\frac{\theta}{2}\right) = \sqrt{\frac{1+e}{1-e}} \tan
\left(\frac{E}{2}\right)
\end{equation}
and Kepler's equation
\begin{equation}
E - e \sin E = M = \frac{2 \pi}{P}(t-T_0).
\end{equation}
where $E$ is the eccentric anomaly and $M$ the mean anomaly.  We used
the \idl--based \rvlin\ package\footnote{{\tt http://exoplanets.org/code/}} \citep{wri09}, which uses a
Levenberg--Marquardt algorithm for the non--linear parts of this
optimization problem.  Orbital parameter uncertainties were estimated
using a bootstrapping technique driven by \rvlin\ (Wang et al. 2012).
We determined the best solution with all parameters
left free, and we also fitted the data
assuming a circular orbit ($e=0$ fixed), with two fewer degrees of freedom.  

We also computed the floating--mean
periodogram (e.g., \citealt{cum99}) for each RV data set and
investigated $\chi^2$ as a function of orbital period over the range
$P$ = 0.1--10,000 d.  We consider the orbital period ``solved'' if (i)
the best--fitting orbit has a $\chi^2$ value at least 10 times smaller
than the second--best alias and (ii) the uncertainty in $P$ for the
best fit is less than 5\%.  Once the first condition was met, we used
a range of initial parameter guesses in RVLIN to be certain we
converged on the correct solution, sometimes changing the inputs only
slightly.

\subsection{Circular or eccentric?}

Determining the eccentricity of long--period sdB binaries is an
astrophysical imperative.  As already mentioned in \S \ref{sec:intro},
if the main sequence companions are responsible for the formation of the
sdB (and the sdB progenitor filled its Roche lobe), one would expect their
orbits to have $e$=0 due to the
circularizing nature of the RLOF and CE processes.  If, however, the hot
subdwarf formed from the merger of a binary originally part of a
triple--star system, the cool companion currently observed would have
had no part in the formation of the hot subdwarf, in which case the
orbit could be non--circular.  All of the long--period systems studied
to date have sufficient room to accommodate an inner binary, even after
allowing for moderate ``adiabatic expansion'' of the orbit
\citep{egg89,deb02} resulting from mass loss in the sdB formation
process.

In order to determine whether eccentric orbit fits are preferred for our
targets, we performed an F--test similar to \citet{luc71} using the statistic
defined by
\begin{equation}
\mathcal{F} = \frac{(\chi^2_{\rm circ} - \chi^2_{\rm ecc})}{(DOF_{\rm
ecc}-DOF_{\rm circ})}\frac{DOF_{\rm ecc}}{\chi^2_{\rm ecc}}
\end{equation}
where $DOF_{\rm ecc}$ and $DOF_{\rm circ}$ are the degrees of freedom in
the eccentric and circular orbit models, respectively, and $\chi^2_{\rm
  ecc}$ and $\chi^2_{\rm circ}$ the chi--squared values of their fits to
the data.  Under the assumption that  $\mathcal{F}$ follows an F--distribution,
this statistic provides a quantitative means for deciding whether we can
justify fitting the data with two additional degrees of freedom ($e$,
$\omega$) compared to our null hypothesis, a circular orbit solution.
As $\mathcal{F}$ increases, so does the likelihood that one can
discredit the null hypothesis.  We reject a
circular orbit solution if the probability of obtaining the observed
$\mathcal{F}$ (the p--value) falls below our false--rejection probability,
which we conservatively set to 10$^{-4}$ due to the considerable phase
gaps in the RV measurements.  If the p--value is above this limit, 
we do not rule out a circular--orbit solution using the current data.

\subsection{Computing the mass ratio}
For those systems for which we can measure the RVs of both components,
we can calculate the mass ratio $q$ of the binary. 
It is straightforward
to show that the instantaneous sdB and main--sequence (MS) companion RVs
at time $t_i$ are related by 
\begin{equation}
V_{\rm MS}(t_i) = \gamma(1+q) - qV_{\rm sdB}(t_i)
\end{equation}
where $q \equiv K_{\rm MS}/K_{\rm sdB} = M_{\rm sdB}/M_{\rm MS}$, and
$\gamma_{\rm sdB} = \gamma_{\rm MS}$ has been assumed. Thus,
$q$ can be measured from the slope of a linear fit to a plot of RV$_{\rm
  MS}$ versus RV$_{\rm sdB}$.  The resulting $q$ is unaffected by any
systematic offsets between the component velocities, such as that caused
by different gravitational redshifts.  It is also independent of the
inclination angle, which remains unknown for all systems studied (none
show eclipses).  We used the {\sc IDL}
\textit{fitexy}\footnote{{\tt http://idlastro.gsfc.nasa.gov/}} function, an
orthogonal distance regression routine (procedure from \citealt{pre92}),
to determine $q$ from the best straight--line fit to the data, taking
into account the errors from both radial velocity measurements.


\section{Results}
\label{sec:results}

All but two systems monitored in our program show some sort of orbital
acceleration; most appear to have periods in excess of 100 days.  Given
such long periods and insufficient phase coverage, we cannot fully
solve for the orbital parameters of the majority of targets
using the 2005--2008 data alone.  Table \ref{tab:preliminary} summarizes
our initial survey results, including the maximum (absolute) RV change
observed for each target, the range of possible orbital periods, and the
best--fitting orbital period taken from the model with the lowest
$\chi^2$ in the periodogram (given in parentheses for the unsolved
systems).  Please note that the solutions for the unsolved systems are
preliminary; they may change significantly with the addition of new
data. The four binaries previously reported as resolved are likely
triple--star systems since their observed accelerations are incompatible
with the linear separation distances implied by their angular separations on the
sky\footnote{Estimates of the minimum projected linear separation distances
 for the candidate triple systems range from 200 to 800 AU.}.  
 Two of these systems show RV variations of the main sequence star
(PG 1040+234, PG 1718+519), while the other two show significant sdB
motions (PG 1253+284, PG 1629+081).  Previous observations of PG
1253+284 showing acceleration of the sdB but not the companion led
\citet{heb02} to a similar triple--star system hypothesis for this
system.  We are currently conducting follow--up observations of unsolved
systems with HET/HRS and will report their orbital parameters in future
publications, as they become available.

\begin{table}
  \caption{Preliminary Survey Results} 
  \label{tab:param1}
  \begin{center}
    \leavevmode
    \begin{tabular}{lcccc} \hline \hline              
Target & No.$^a$&RV & Period & $\Delta RV_{\rm max}$\\
             & Obs.  &      variable?           & [days]  &  [km s$^{-1}$]\\
\hline
\multicolumn{5}{c}{\textit{Unsolved Systems}}\\
\hline
PG 0039+049 &  7& yes$^b$ &  150-300 (220) & 11 \\
PG 0110+262&	 9 & ? & ?& --\\
PB 8783 & 8 & ?& ? & --\\
PHL 1079& 5 & yes$^b$ & 80--500 (122)& 24\\
PG 0232+095&	6& yes$^b$ & $<$200 (14)&29\\	
PG 1040+234&	 9& yes$^b$ & $>$500 (700)&12\\
PG 1206+165& 6 & yes$^b$ & ? & 15\\
PG 1253+284& 11& yes$^c$ &  $<$220 (161)& 93\\  
PG 1449+653&	  8& yes$^b$ & 750--1000 (859)& 15\\  
PG 1629+081& 8 & yes$^c$ & 2--11 (2.9) & 107 \\   
PG 1701+359&	14& yes$^b$ & 630--830 (725) & 9 \\  
PG 1718+519& 8	& yes$^b$ & 200--330 (300)&7\\
\hline		
\multicolumn{5}{c}{\textit{Solved Systems}}\\
\hline
PG 1104+243 & 10 & yes$^{b,c}$ & 753.2 & 16\\
PG 1317+123 & 11 & yes$^{b,c}$ & 1179 & 13\\
PG 1338+611 & 11 & yes$^{b,c}$ & 937.5& 16 \\
\hline
\multicolumn{5}{l}{$^a$number of useful HET/MRS observations from 2005-2008}\\
\multicolumn{5}{l}{$^b$cool companion acceleration measured.}\\
\multicolumn{5}{l}{$^c$hot subdwarf acceleration measured.}

    \end{tabular}
  \end{center}
  \label{tab:preliminary}
\end{table}

 \begin{table*}
  \caption{Cool Companion Velocities for Solved Systems} 
  \label{tab:param2}
  \begin{center}
\scriptsize
    \leavevmode
    \begin{tabular}{cclcclccl} 
    \hline \hline              
HJD& RV$_{\rm MS}$ & Facility$^a$ &HJD& RV$_{\rm MS}$ & Facility$^a$ &HJD&
RV$_{\rm
MS}$ & Facility$^a$ \\
$-$2450000 & [km s$^{-1}$] &   & $-$2450000 & [km s$^{-1}$] &   &$-$2450000 &
[km s$^{-1}$] &   \\
\hline
\multicolumn{3}{c}{PG 1104+243} & \multicolumn{3}{c}{PG 1317+123} &
\multicolumn{3}{c}{PG 1338+611} \\
\multicolumn{3}{c}{---------------------------------------------} &
\multicolumn{3}{c}{---------------------------------------------} &
\multicolumn{3}{c}{---------------------------------------------} \\
     -215.23370 &   -16.80 $\pm$  4.70 & K &      3379.93771 &    45.66 $\pm$
1.09 & H &      3379.95036 &    24.82 $\pm$  0.96 & H \\
     -214.16710 &   -19.30 $\pm$  4.50 & K &      3392.92593 &    44.28 $\pm$
1.00 & H &      3400.89553 &    25.62 $\pm$  1.21 & H \\
     -213.24680 &   -17.30 $\pm$  4.70 & K &      3421.84076 &    44.04 $\pm$
0.68 & H &      3415.89117 &    27.84 $\pm$  0.77 & H \\
     -212.32400 &   -17.60 $\pm$  4.40 & K &      3446.76922 &    44.01 $\pm$
0.63 & H &      3429.87462 &    28.35 $\pm$  0.57 & H \\
     3379.84221 &   -20.53 $\pm$  0.74 & H &      3476.68120 &    42.92 $\pm$
0.92 & H &      3476.70485 &    31.35 $\pm$  0.63 & H \\
     3392.80703 &   -20.20 $\pm$  0.72 & H &      3488.65464 &    42.87 $\pm$
0.91 & H &      3488.66280 &    32.69 $\pm$  0.96 & H \\
     3415.95642 &   -20.06 $\pm$  0.56 & H &      3500.62543 &    42.13 $\pm$
0.70 & H &      3502.64537 &    32.77 $\pm$  0.64 & H \\
     3429.73480 &   -19.63 $\pm$  0.41 & H &      3724.01307 &    36.39 $\pm$
0.51 & H &      3731.99294 &    40.23 $\pm$  0.78 & H \\
     3447.65376 &   -18.90 $\pm$  0.54 & H &      3771.87931 &    35.42 $\pm$
0.78 & H &      3732.97699 &    40.65 $\pm$  0.73 & H \\
     3462.62302 &   -19.61 $\pm$  0.40 & H &      4453.00102 &    48.53 $\pm$
1.23 & H &      4479.96424 &    35.36 $\pm$  0.78 & H \\
     3476.78809 &   -19.43 $\pm$  0.51 & H &      4551.75800 &    45.21 $\pm$
0.90 & H &      4540.80436 &    37.03 $\pm$  0.83 & H \\
     3498.74435 &   -19.12 $\pm$  0.46 & H &      5030.42431 &    34.08 $\pm$
0.09 & M &      5028.44884 &    27.83 $\pm$  0.02 & M \\
     3757.80663 &   -11.03 $\pm$  0.40 & H &      5030.42431 &    34.08 $\pm$
0.09 & M &      5028.47026 &    27.77 $\pm$  0.02 & M \\
     4479.82452 &   -10.97 $\pm$  0.78 & H &      5030.45097 &    32.69 $\pm$
0.17 & M &      5028.49174 &    27.77 $\pm$  0.03 & M \\
     5204.70741 &   -12.24 $\pm$  0.09 & M &      5030.45097 &    32.69 $\pm$
0.17 & M &      5341.57223 &    30.81 $\pm$  0.03 & M \\
     5204.72527 &   -12.37 $\pm$  0.04 & M &      5294.54985 &    39.10 $\pm$
0.10 & M &      5351.41840 &    31.07 $\pm$  0.05 & M \\
     5217.67376 &   -12.29 $\pm$  0.10 & M &      5294.54985 &    39.10 $\pm$
0.10 & M &      5371.41748 &    32.84 $\pm$  0.02 & M \\
     5217.69174 &   -12.27 $\pm$  0.10 & M &      5353.40135 &    39.53 $\pm$ 
0.11 & M &      5566.78096 &    40.00 $\pm$  0.04 & M \\
     5234.60506 &   -11.42 $\pm$  0.04 & M &      5353.40135 &    39.53 $\pm$ 
0.11 & M &      5616.64021 &    40.29 $\pm$  0.02 & M \\
     5234.61608 &   -12.09 $\pm$  0.11 & M &      5373.41275 &    41.12 $\pm$ 
0.08 & M &      5616.64021 &    40.29 $\pm$  0.02 & M \\
     5234.62710 &   -12.02 $\pm$  0.12 & M &      5373.41275 &    41.12 $\pm$ 
0.08 & M &      5622.70741 &    39.98 $\pm$  0.04 & M \\
     5264.54423 &   -11.15 $\pm$  0.07 & M &      5566.74478 &    46.66 $\pm$ 
0.13 & M &      5622.73169 &    39.96 $\pm$  0.03 & M \\
     5340.42821 &   -11.93 $\pm$  0.07 & M &      5566.74478 &    46.66 $\pm$ 
0.13 & M &      5656.60318 &    39.64 $\pm$  0.04 & M \\
     5351.37760 &   -11.91 $\pm$  0.08 & M &      5616.61438 &    47.49 $\pm$ 
0.06 & M &      5658.58702 &    39.98 $\pm$  0.08 & M \\
     5553.70059 &   -18.20 $\pm$  0.11 & M &      5616.61438 &    47.49 $\pm$ 
0.06 & M &      5660.61069 &    39.86 $\pm$  0.04 & M \\
     5569.69768 &   -18.78 $\pm$  0.09 & M &      5616.61438 &    47.49 $\pm$ 
0.06 & M &      5663.54111 &    39.90 $\pm$  0.03 & M \\
     5579.56553 &   -19.22 $\pm$  0.12 & M &      5616.61438 &    47.49 $\pm$ 
0.06 & M &      5666.53453 &    39.63 $\pm$  0.02 & M \\
     5579.58348 &   -18.93 $\pm$  0.09 & M &      5640.61157 &    45.88 $\pm$ 
0.06 & M &         &    &  \\
     5589.75559 &   -19.06 $\pm$  0.10 & M &      5640.61157 &    45.88 $\pm$ 
0.06 & M &         &     &  \\
     5611.64081 &   -19.73 $\pm$  0.07 & M &      5653.60091 &    44.58 $\pm$ 
0.11 & M &         & &  \\
     5622.61268 &   -19.95 $\pm$  0.08 & M &      5653.60091 &    44.58 $\pm$ 
0.11 & M &         &  &  \\
     5639.58778 &   -19.57 $\pm$  0.07 & M &      5661.63454 &    45.09 $\pm$ 
0.09 & M &         &  &  \\
     5650.50737 &   -20.16 $\pm$  0.07 & M &      5661.63454 &    45.09 $\pm$ 
0.09 & M &         &    & \\
     5652.55550 &   -19.99 $\pm$  0.05 & M &      5722.45431 &    44.93 $\pm$ 
0.24 & M &         &   &  \\
     5655.60537 &   -19.87 $\pm$  0.07 & M &      5722.45431 &    44.93 $\pm$ 
0.24 & M &         &    &  \\
     5658.47913 &   -20.10 $\pm$  0.06 & M &         &     &  &         &    & 
\\
     5664.51354 &   -20.17 $\pm$  0.07 & M &         &    &  &         &   & 
\\
     5666.43280 &   -20.41 $\pm$  0.05 & M &         &     & &         &    & 
\\
     5666.45153 &   -20.32 $\pm$  0.06 & M &         &     &  &         &     &
 \\
     5672.53688 &   -20.30 $\pm$  0.11 & M &         &   &  &         &     & 
\\
     5685.44968 &   -20.02 $\pm$  0.06 & M &         &   &  &         &   &  \\
     5718.44565 &   -19.50 $\pm$  0.06 & M &         &    &  &         &   & \\
\hline
\multicolumn{9}{l}{$^a$K = KPNO 2.1m/GoldCam spectrograph \citep{oro97},
H=HET/MRS, M=Mercator/HERMES (\ostrv ).}
    \end{tabular}
  \end{center}
  \label{tab:rvs}
\end{table*}

 \begin{table*}
  \caption{Hot Subdwarf Velocities for Solved Systems from HET/MRS} 
  \label{tab:param3}
  \begin{center}
    \leavevmode
    \begin{tabular}{cccccc} 
    \hline \hline              
HJD& RV$_{\rm MS}$  &HJD& RV$_{\rm MS}$ &HJD& RV$_{\rm MS}$  \\
$-$2450000 & [km s$^{-1}$] &    $-$2450000 & [km s$^{-1}$]   &$-$2450000 & [km
s$^{-1}$]   \\
\hline
\multicolumn{2}{c}{PG 1104+243} & \multicolumn{2}{c}{PG 1317+123} &
\multicolumn{2}{c}{PG 1338+611} \\
\multicolumn{2}{c}{---------------------------------------------} &
\multicolumn{2}{c}{---------------------------------------------} &
\multicolumn{2}{c}{---------------------------------------------} \\
   3379.84221 &     -7.8 $\pm$   1.9  &      3379.93771 &     28.8 $\pm$   1.3
&      3379.95036 &     41.8 $\pm$   3.6\\
     3392.80703 &     -8.0 $\pm$   1.6  &      3392.92593 &     31.9 $\pm$  
2.6 &      3400.89553 &     43.3 $\pm$   3.3\\
     3415.95642 &     -9.6 $\pm$   2.8  &      3421.84076 &     31.4 $\pm$  
1.0 &      3415.89117 &     39.1 $\pm$   2.9\\
     3429.73480 &    -11.1 $\pm$   2.3  &      3446.76922 &     32.2 $\pm$  
2.2 &      3429.87462 &     38.3 $\pm$   3.6\\
     3447.65376 &    -11.3 $\pm$   1.1  &      3476.68120 &     31.8 $\pm$  
1.0 &      3476.70485 &     32.5 $\pm$   2.4\\
     3462.62302 &    -11.4 $\pm$   2.0  &      3488.65464 &     34.6 $\pm$  
3.0 &      3488.66280 &     31.4 $\pm$   3.5\\
     3476.78809 &    -12.0 $\pm$   2.3  &      3500.62543 &     37.6 $\pm$  
2.9 &      3502.64537 &     33.1 $\pm$   4.4\\
     3498.74435 &    -11.8 $\pm$   2.4  &      3724.01307 &     50.1 $\pm$  
1.8 &      3731.99294 &     18.4 $\pm$   3.6\\
     3757.88245 &    -22.5 $\pm$   1.5  &      3771.87931 &     51.3 $\pm$  
5.4 &      3732.97699 &     16.0 $\pm$   2.1\\
     4479.87074 &    -23.6 $\pm$   2.3  &      4453.00102 &     24.2 $\pm$  
8.3 &      4479.96424 &     28.3 $\pm$   3.9\\
&     &      4551.75800 &     31.2 $\pm$   3.0 &      4540.80436 &     20.1
$\pm$   6.2\\

\hline
    \end{tabular}
  \end{center}
  \label{tab:rvs_sdB}
\end{table*}

\begin{table*}
  \caption{Orbital Parameters for the Solved Systems} 
  \label{tab:param4}
\footnotesize
  \begin{center}
    \begin{tabular}{crlrlrlrlrlrlrlrl} \hline \hline              
Target & \multicolumn{2}{c}{$P$} & \multicolumn{2}{c}{$K_{\rm MS}$} &
\multicolumn{2}{c}{$K_{\rm sdB}$$^{a}$} &\multicolumn{2}{c}{$\gamma$}
&\multicolumn{2}{c}{$e$}  & \multicolumn{2}{c}{$\omega$}  &
\multicolumn{2}{c}{$T_{\rm 0}$$^{c}$} \\
 Name& \multicolumn{2}{c}{[days] }& \multicolumn{2}{c}{[km s$^{-1}$]} &
\multicolumn{2}{c}{[km s$^{-1}$]} &\multicolumn{2}{c}{[km s$^{-1}$]}
&\multicolumn{2}{c}{} &\multicolumn{2}{c}{[deg]} 
&\multicolumn{2}{c}{[HJD$-$2450000]} \\
\hline
PG 1104+243 & 753.2& \hspace{-2.5mm}$\pm$ 0.8 & 4.43&\hspace{-2.5mm}$\pm$ 0.06
& 6.5 &\hspace{-2.5mm}$\pm$ 0.8 &$-$15.68 &\hspace{-2.5mm}$\pm$ 0.05 &
\multicolumn{2}{c}{0.0$^{b}$}  &    \multicolumn{2}{c}{--}  &   
4534.5 &\hspace{-2.5mm}$\pm$ 3.3  \\   

PG 1317+123 &  1179 &\hspace{-2.5mm}$\pm$ 12 & 6.2&\hspace{-2.5mm}$\pm$ 0.2 &
15.5 &\hspace{-2.5mm}$\pm$ 1.7 &+40.3 &\hspace{-2.5mm}$\pm$ 0.2 &
\multicolumn{2}{c}{0.0$^{b}$} & \multicolumn{2}{c}{--}  &   
4453  &\hspace{-2.5mm}$\pm$ 15  \\

PG 1338+611 & 937.5 &\hspace{-2.5mm}$\pm$ 1.1 & 8.8& \hspace{-2.5mm}$\pm$  0.15
& 15.2 &\hspace{-2.5mm}$\pm$ 1.8 &+32.58 &\hspace{-2.5mm}$\pm$ 0.07 & 0.15
&\hspace{-2.5mm}$\pm$ 0.02 & 209 & \hspace{-2.5mm}$\pm$ 6 &   
4685.5 &\hspace{-2.5mm}$\pm$ 3.0    \\
\hline
\multicolumn{15}{l}{$^{a}$computed from $K_{\rm MS}$ and $q$.}\\
\multicolumn{15}{l}{$^{b}$fixed; no reason to prefer eccentric fit over
circular
solution at this time.}\\
\multicolumn{15}{l}{$^{c}$time of velocity maximum (maximum redshift) in the
cool companion RV curve ($e=0$ case) or time of periastron passage ($e \ne
0$).}\\
    \end{tabular}
  \end{center}
  \label{tab:system_parameters}
\end{table*}

\ostrv\ began a similar RV monitoring program with Mercator/HERMES in
2009, shortly after the HET observations reported here ended.  Three of
their targets (PG 1104+243, PG 1317+123, PG 1338+611) overlap with the
HET sample.  By phasing together both observation sets, we are able to
improve upon their original system constraints significantly.  In the
case of PG 1104+243, we also include RVs obtained in 1995 with the
GoldCam spectrograph on the 2.1m telescope at Kitt Peak National
Observatory \citep{oro97}.  We discuss all three systems and their
updated orbital parameters below.  Tables \ref{tab:rvs} and
\ref{tab:rvs_sdB} present the heliocentric RVs of the cool companion and
hot subdwarf, respectively, while Table \ref{tab:system_parameters}
gives the fitted system parameters described in \S 4.3 along with their formal errors
from the multi--parameter fits.  

The systemic velocity ($\gamma$) is subject to a possible zero--point
offset, so although we quote only the formal errors in the table, we
refer the reader back to the discussion of our zero--point checks in \S
\ref{subsec:measuring_RVs} for a more complete understanding of their
likely uncertainties.  Due to the low quality of the hot subdwarf RVs,
the orbital parameters shown ($P$, $K_{\rm MS}$, $\gamma$, $e$,
$\omega$, $T_{\rm 0}$) were derived from the cool companion measurements
alone.  The $K_{\rm sdB}$ values reported were calculated from $K_{\rm
  MS}$ and $q$, as described in \S 4.5; in all three cases, they are
consistent with direct orbit fits to the hot subdwarf RVs if $P$ is
taken from the cool companion orbital fit and the relation $\omega_{\rm
  sdB} = \omega_{\rm MS}+\pi$ is imposed.  Finally, we detected no
significant velocity offsets between telescopes, so we fixed their
relative zero--point offsets to zero during the final orbit fitting.

  \begin{figure*}
  \begin{center}
   \includegraphics[scale=0.99]{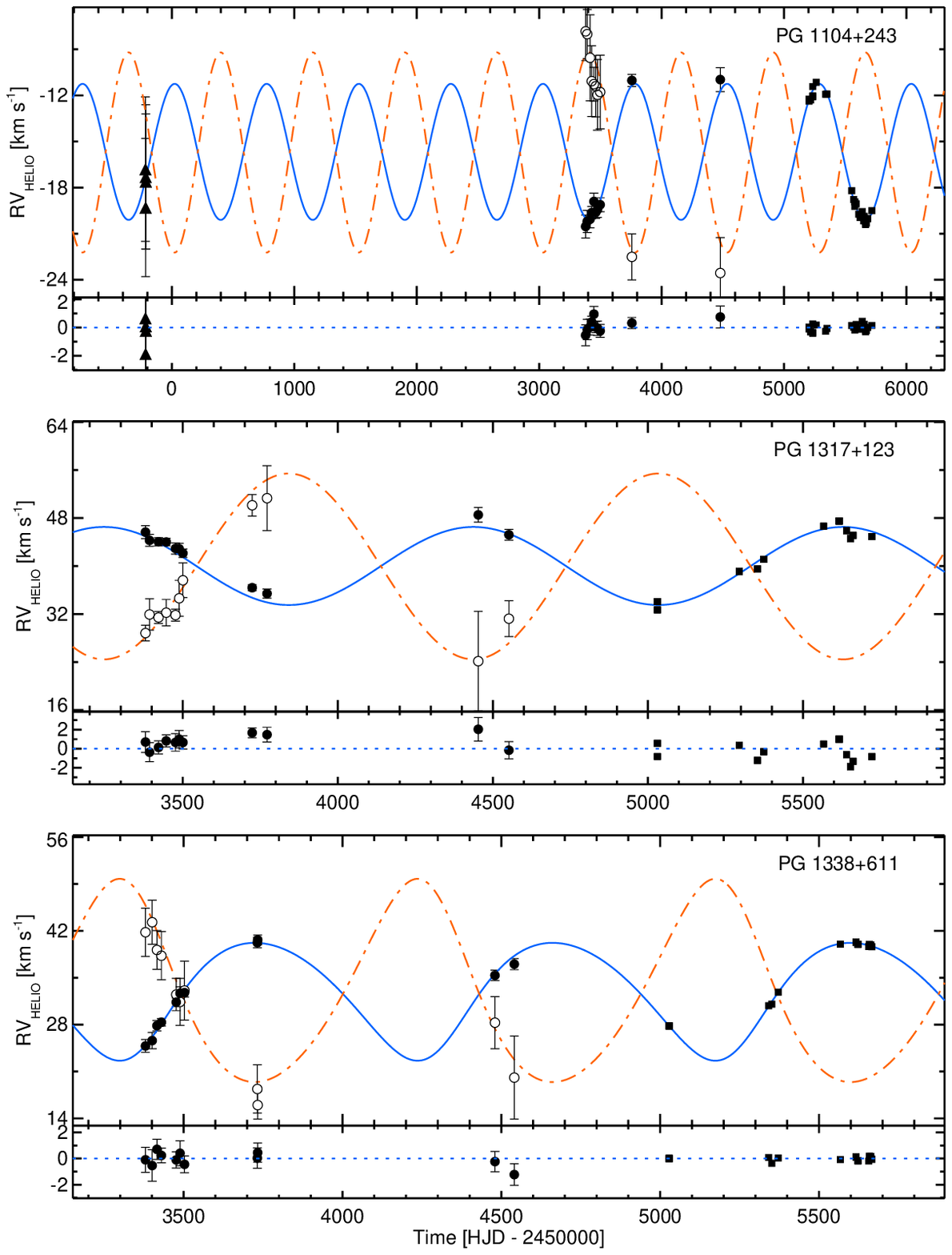}
   \caption{Heliocentric RV curves for the three targets
     overlapping with the study of \ostrv: PG 1104+243 (\textit{top}),
     PG 1317+123 (\textit{middle}), and PG 1338+611 (\textit{bottom}).
     Cool companion measurements from HET/MRS, \ostrv, and \citet{oro97}
     are shown with filled circles, squares, and triangles,
     respectively.  Open circles represent sdB velocities measured from
     He lines in the HET/MRS spectra.  The best--fitting orbital
     solutions for the sdB and cool companion are denoted with
     dot--dashed (orange) and solid (blue) lines, respectively.
     Residuals from this fit are shown in the lower portion of each
     panel for the cool companion measurements.  Error bars on the
     Mercator/HERMES points are much smaller than the symbol sizes;
     we do not plot them here.}
     \label{fig:rv}
  \end{center}
\end{figure*}

 \begin{figure*}
  \begin{center}
   \includegraphics{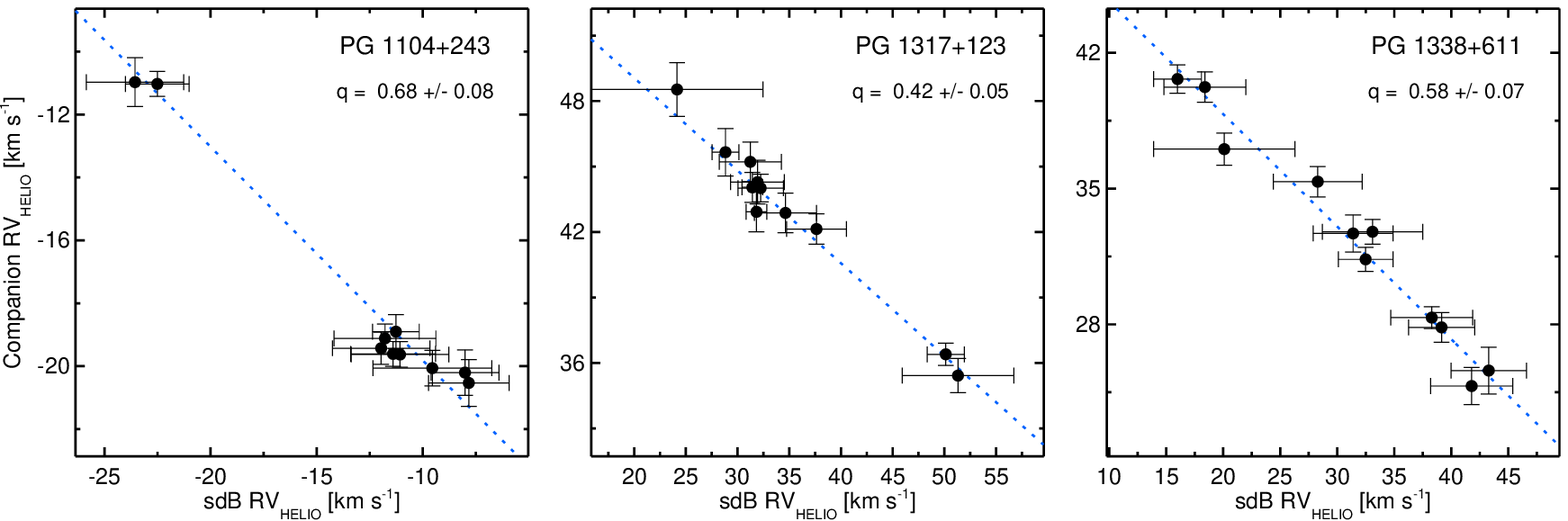}
   \caption{HET/MRS orbital velocities of the cool companion and sdB
     plotted against one another for PG 1104+243, PG 1317+123, and PG
     1338+611.  The dotted lines show the best--fitting linear functions
     to the data; the absolute values of their slopes are the mass
     ratios.  Improved constraints on the mass ratios from HERMES
     measurements of the sdB velocities will be presented by Vos et al.
     (2012, in prep).}
     \label{fig:q}
  \end{center}
\end{figure*}

\subsection{PG 1104+243}
Previous classifications for the cool companion of PG 1104+243 have
ranged from F8\,V \citep{staphd} to K3.5\,V \citep{all94}.  Comparisons
of the HET/MRS spectra to the ELODIE templates show a best--fitting
spectral type near G2\,V.  The dilution of the companion spectrum
is $D \equiv L_{\rm comp}/L_{\rm total} \approx 0.5$,
where the luminosity ratio refers to the 5075--5200 \AA\ spectral region.
As noted by \ostrv, the hot subdwarf in PG
1104+243 might be described best as an `sdOB' star since its spectrum
shows the \ion{He}{2} 4686 \AA\ line in addition to strong H Balmer and
\ion{He}{1} features.  Previous RV measurements have been
published by \citet{oro97}, who found no significant variations over a
three--day timespan.  An orbital solution with $P$ = 752 $\pm$ 14 d was
recently published by \ostrv.  Combining our data with those of both
\ostrv\ and \citet{oro97}, we find an orbital period of $P$ = 758.8
$\pm$ 0.13 d, in excellent agreement with the \ostrv\ result.  This
value falls frustratingly close to two years, as clearly seen in the RV
curve shown in Figure \ref{fig:rv}.  By chance, current observing seasons
align with quadrature phases; otherwise, all points would have been
obtained near a conjunction, a less--than--ideal situation for measuring
the mass ratio from spectroscopy.  At the same time, information near
conjunction is needed to improve constraints on the geometry of the
orbit.  Several more orbital cycles must elapse before complete
phase coverage can be obtained.  At this time, we have no reason to
prefer an eccentric orbit over a circular one.

The presence of strong He lines from the sdB (5876 \& 4686 \AA) permits
us to measure accelerations for both components.  As expected for stars
in a binary, the sdB and MS velocity curves are 180 degrees out of
phase.  The sdB RV curve shows a systemic velocity consistent with that
measured from the companion ($\gamma = -15.5 \pm 0.03$ km s$^{-1}$).
From the relationship between RV$_{\rm sdB}$ and RV$_{\rm MS}$, as seen
in Figure \ref{fig:q}, we compute a mass ratio of $q$ = 0.68 $\pm$ 0.08.
Table \ref{tab:system_parameters} summarizes the system parameters
derived for PG 1104+243. If we assume a companion mass of 1.0 M$_{\sun}$
and adopt a 10\% error to include uncertainties in the spectral
classification, we find M$_{\rm sdB}$ = 0.68 $\pm$ 0.11 M$_{\sun}$,
considerably larger than the `canonical' value of $\sim$0.48 M$_{\sun}$,
which corresponds to helium core ignition at the tip of the RGB.  If we
instead assume exactly the canonical mass for the hot subdwarf, we would
compute a mass of 0.7 $\pm$ 0.1 M$_{\sun}$ for the companion, which
implies its spectral type might be significantly later than our result
from the composite spectrum analysis, or that the companion is evolved.

\subsection{PG 1317+123 (Feige 80)}

Spectral classifications for the hot component in PG 1317+123 have
varied over years.  It is probably an sdO star as it displays a strong
\ion{He}{2} 4686 \AA\ line and a relatively weak \ion{He}{1} 4472 \AA\
line.  \citet{ull98} classify the cool companion as a G8\,V, while
\ostrv\ report it to be a `G' star.  We find a spectral type near G1\,V
and dilution $D \approx 0.3$ from our comparisons with ELODIE models.
\ostrv\ reported an orbital period of $P$ = 912 $\pm$ 52 d.  After
phasing the HET data with theirs, we find a significantly longer value,
$P$ = 1179 $\pm$ 12 d.  Figure \ref{fig:rv} shows the resulting RV
curve, while Table \ref{tab:system_parameters} summarizes all of the
orbital parameters derived.  Our results for the systemic velocity and
MS RV amplitudes also disagree. As pointed out by \ostrv, the
Mercator/HERMES measurements show much larger scatter than one would
expect given the precision of their data.  They briefly suggested
$\gamma$ Doradus--type pulsations or star spots on the cool companion as
potential culprits.  However, G--type classifications for the cool
companion are inconsistent with those of $\gamma$ Dor stars, which are
typically early F or late A dwarfs.  The other scenario, a bright spot
on the companion's surface, seems more compatible with the observations.
Assuming a solar--type dwarf and a star spot with a filling factor of a
few percent \citep{saa97}, the amplitude of the scatter (1--2 km
s$^{-1}$) would imply a rotational period around one week.  As with PG
1104+243, we currently have no reason to prefer an eccentric--orbit
solution.  Even with additional data, constraining the eccentricity will
be especially difficult until the rapid velocity variations are better
characterized.

\ion{He}{1} 5876 \AA\ and \ion{He}{2} 4686 \AA\ features provided RV
estimates for the hot subdwarf that are again 180 degrees out of phase
with the cool companion.  Figure \ref{fig:q} shows a strong linear
relationship between RV$_{\rm sdO}$ and RV$_{\rm MS}$, whose slope gives
$q$ = 0.42 $\pm$ 0.05.  A circular orbit fitted to the sdO RVs with the
period and phase fixed to the values in Table
\ref{tab:system_parameters} gives a consistent systemic velocity and a
semi--amplitude in agreement with the expectation from combining $K_{\rm
  MS}$ and $q$.  Assuming a companion mass from with our spectral
classification, $M_{\rm MS}$ = 1.03 $\pm$ 0.11 M$_{\sun}$, gives M$_{\rm
  sdO}$ = 0.43 $\pm$ 0.07 M$_{\sun}$.

\subsection{PG 1338+061 (Feige 87)}

\begin{figure}
  \begin{center}
   \includegraphics[scale=1.01]{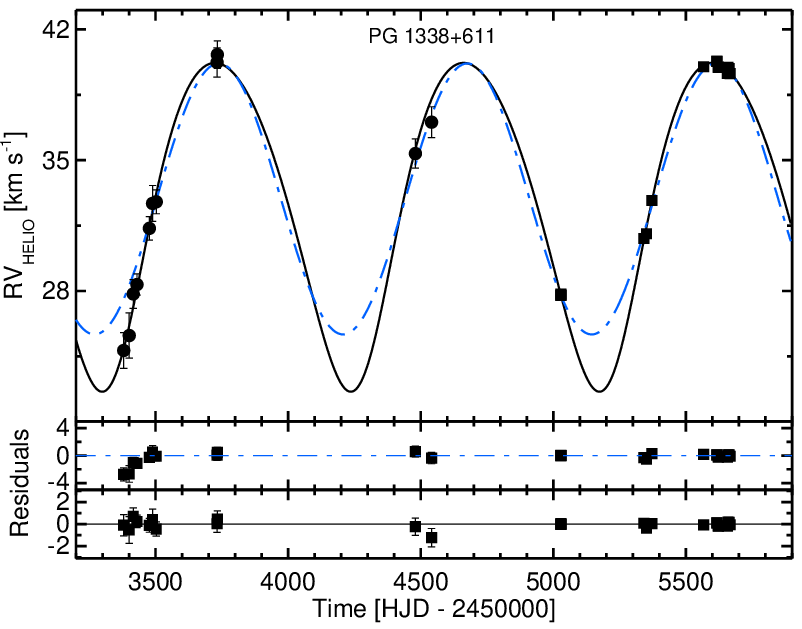}
   \caption{Heliocentric radial velocities for PG 1338+611 with the
     best--fitting circular (blue dotted line) and eccentric (solid black 
     line) orbits, and their residuals (bottom panels).  Velocities from
     HET/MRS and \ostrv\ are shown with circles and squares,
     respectively.  The data are not consistent with a circular
     solution; the best--fitting eccentric model has $e$ = 0.15 $\pm$
     0.02.}
     \label{fig:1338}
  \end{center}
\end{figure}

Our analysis of PG 1338+061's spectrum reveals a G2--G7\,V companion
(best fit: G4\,V with $D \approx 0.2$), much earlier than previous
classifications (K0\,V, \citealt{staphd}; K4.5\,V, \citealt{all94}).
Two \ion{He}{1} lines (5876 \& 4472 \AA) were used to measure the sdB
RVs, which show the sdB and companion accelerating in anti--phase with
one another.  The last panel in Figure \ref{fig:rv} presents the RV
curve for PG 1338+061, combining the HET and Mercator observations.
Unlike PG 1104+243 and PG 1317+123, we find significant evidence for an
orbit with non--zero eccentricity.  Figure \ref{fig:1338} shows our
best--fitting circular and eccentric fits to the cool companion's RV
curve for comparison.  Computation of the F--test statistic gives
$\mathcal{F}$=12.6 with a corresponding p--value of $10^{-7}$, well
below our criterion for claiming non--zero eccentricity.  We therefore
reject the circular--orbit hypothesis and report $e$ = 0.15 $\pm$ 0.02 for
this system.  The best--fitting period in this case is $P = 937 \pm 10$
d, in slight disagreement with \ostrv, who assumed a circular--orbit
solution and found $P$ = 915 $\pm$ 16 d.  In \S \ref{subsec:e} we
discuss the implications of non--zero eccentricity on formation scenarios of
long--period sdB systems.

The relationship between $RV_{\rm sdB}$ and $RV_{\rm MS}$ (Figure
\ref{fig:q}) is fitted well with $q = 0.58 \pm 0.07$.  Combining this
value with $K_{\rm MS}$ predicts a velocity semi--amplitude consistent
with the best--fit to the subdwarf RV curve (with period and phase
fixed).  
 The difference in the systemic velocities is
$\gamma_{\rm sdB} - \gamma_{\rm MS} = - 2.1 \pm$ 1.0 km s$^{-1}$.
Assuming a mass of $M_{\rm MS}$ = 0.97 $\pm$ 0.1 $M_{\sun}$ for the
companion, we calculate $M_{\rm sdB}$ = 0.56 $\pm$ 0.09 $M_{\sun}$ for
the hot subdwarf, consistent with the canonical value.

\vspace{6mm}
 \section{The Period Distribution of \\Hot Subdwarf Binaries Revisited}
 \label{sec:histogram}
 
   \begin{figure}
  \begin{center}
   \includegraphics[scale=1.]{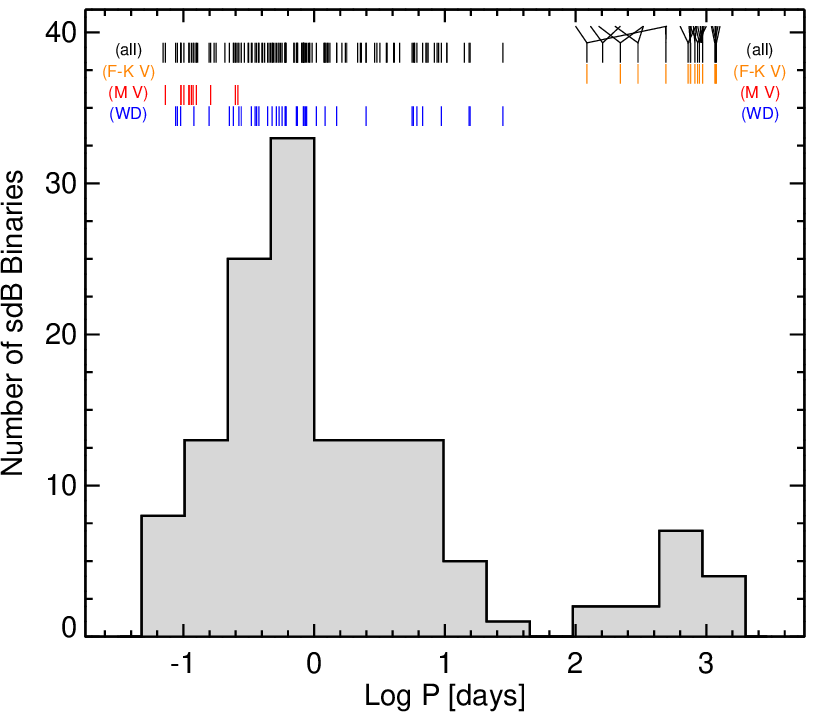}
   \caption{Histogram of the orbital periods of 140 hot subdwarf
     binaries with known orbital parameters.  The black vertical lines
     at the top indicate the individual periods of all systems.  For the
     long--period binaries with poor fits, they represent the
     best--fitting orbit in the periodogram; two diagonal lines
     emanating from the tops of these symbols mark the range of possible
     orbital periods.  In the rows of vertical lines below the first, we
     re--mark systems with F--K dwarf (second row, orange marks), M
     dwarf (third row, red marks), and white dwarf (fourth row, blue
     marks) companions.  Those with uncertain companion classifications
     or none at all are not shown. The long--period systems exclusively
     have companions of spectral type $\sim$K4 or earlier.  Although the
     statistics are poor, observations currently show a gap centered
     near $P$ = 70 d.}
     \label{fig:period_histogram}
  \end{center}
\end{figure}
 

Recent orbital solutions for long--period binaries have drastically
changed the empirical period distribution of hot subdwarf binaries,
compared with the situation a few years ago.  We present in Figure
\ref{fig:period_histogram} an updated period histogram for all systems
with published orbital parameters.  Most of the short--period systems can
be found in \citet{muchfuss} (see Table A.1 and references therein), 
\citet{cop11}, and \citet{mor03}.  For the long--period binaries, which
include PG 1018$-$047 \citep{dec11}, results from \ostrv, and all
targets in our survey with reasonably well--constrained periods (see
Table \ref{tab:preliminary}), we include estimates of the uncertainties
in the orbital periods.  Given the difficulties associated with
monitoring the long--period systems, several strong observational biases are
undoubtedly at play in the diagram; premature conclusions should not be
drawn.

\vspace{10mm}
\subsection{The problem posed by long orbital periods}
\label{subsec:long_period_problems}
One clear trend appears to emerge from the empirical distribution of
orbital periods: closer binaries tend to have late--type MS or WD
companions while the long--period systems host F/G/K dwarfs. This
apparent dichotomy has been noted before \citep[e.g.,][]{gre08}. Although
the number of long--period binaries known is still relatively small,
their addition allows us to begin testing the predictions of BPS models
in {\em both} the short-- and long--period regimes for the first time.

The \citet{han02,han03} models reproduce the short--period end 
of the distribution quite well but predict that the sdB+MS (G/K) binaries 
should have periods shorter than $\sim$100 d, the outcome of CE evolution 
using the ``$\alpha$--formalism'', in which orbital energy (and/or thermal energy 
stored in the envelope) is tapped to expel the envelope. Current observations 
suggest that the energy--based $\alpha$--formalism for CE evolution 
in its simplest forms does not
account for the long--period sdB binaries. We note that \citet{han02,han03} 
deliberately excluded the ÕGK--effectÕ binaries when choosing their 
favored BPS models, focusing on the short--period M dwarf and WD systems
(the only systems with measured periods available at the time). Thus, it is 
not surprising their model predictions are inconsistent with the current period 
distribution of sdB+F/G/KV binaries. \citet{cla12} encountered a similar problem 
when modeling the sdB population using the $\alpha$--formalism.
This failing is not limited to the sdB binaries and has been discussed in
other contexts by \citet{web08}, who points out the need in some cases
for (1) additional energy sources to augment or substitute for some of
the energy extracted from the orbit in expelling the envelope, or (2) a
reduction in the binding energy of the envelope (possibly through more
extensive mass loss prior to the CE phase), in order that orbital
shrinkage not be so drastic.

An alternative scenario proposed by \citet{nel00,nel01a} appears, at
first glance, to be more successful.  In this model, the long--period
systems are the product of CE evolution, with a prescribed fractional
angular momentum loss that is proportional to the fractional mass lost
when the envelope is ejected.  The current distribution of long--period
systems overlaps surprisingly well with the densest region of the
predicted population when the angular momentum parameter $\gamma = 1.5$
(Figure 2 of \citealt{nel10}).  Current observations hint at a dearth of
binaries in the range $P$ $\approx$ 30--110 d, and in this context we
note that a period histogram drawn `by eye' from the sdB+MS population
(with companion masses $\sim 1\,M_{\sun}$) predicted by \citet{nel10}
shows a local minimum in the number of binaries with periods from
$\approx$ 30 to 90 d.  It would be well not to read too much into this
apparent agreement, since the model is advertised as tentative, is not
well--matched to the PG sample in terms of its flux limit, and the
transformation of sdB bolometric luminosities to $V$ magnitudes is not
described.  In addition, the empirical diagram should be more thoroughly
populated and the observational biases accounted for, before the
significance of the gap is assessed.

\vspace{8mm}
\subsection{Eccentricity as a clue to the long periods}
\label{subsec:e}
It is interesting to consider the question of eccentricity of the
observed wide orbits.  PG 1338+061 (this paper) and possibly PG
1018$-$047 \citep{dec11} have significant eccentricities.  We also note
that eccentric orbits have not been excluded for PG 1104+243 and PG
1317+123 based on the present data.  If wide sdB binaries typically
prove to have eccentric orbits, this would suggest one of three
possibilities. Perhaps (1) these systems are or were hierarchical triple
systems in which the sdB star was made in the inner binary, possibly by
merger (see, e.g., \citealt{cla11}, who describe a ``hydrogen merger
channel'' that leads to sdB's with canonical masses). 
We note that in PG 1338+061, the minimum periastron distance
of 2 AU in the current orbit gives ample room to have once accommodated
an inner system.  If however the systems have always been binaries,
then either (2) circularization was not achieved because
 the sdB progenitor never filled its Roche lobe or (3) eccentricity
 was pumped into the system via one of several possible mechanisms. 

In the first case, the current long--period eccentric orbits are not
directly comparable to the predictions of BPS models since the observed
cool companions did not participate directly in the formation of the hot
subwarfs.  They nevertheless provide some dynamical constraints on
``binary'' formation scenarios, since they limit the size of the inner
orbit in the hierarchical triple.  Studies of the sdB rotational
velocities in these systems might shed light on whether a merger
included two He--core WDs (fast rotation expected: \citealt{gou06};
although if there is a giant stage between merger and the sdB phase the
rotation may be slowed, see \citealt{zha12}) or a He WD and a MS star
(slow rotation, owing to angular momentum shedding near the tip of the
RGB; \citealt{cla11}).

In the case these systems have always been binaries, we suppose that a strong, tidally--enhanced wind
along the lines suggested by \citet{tou88} can occasionally result in
the complete loss of the envelope, just as the core of the sdB
progenitor reaches the critical mass (or near it, see \citealt{dcr96})
to ignite He burning, without ever filling its Roche lobe. Some
expansion of the orbit would occur owing to the wind mass loss, and the
orbital eccentricity could in large part be preserved. If this timing
coincidence does not occur, the extra mass loss can at least reduce the
envelope binding energy and, after a CE episode, leave a (circular)
orbit at longer period than would otherwise be attained, or even reduce
the mass ratio to the point where RLOF mass transfer proceeds stably and
a CE phase is avoided altogether, as in \citet{tou88}.

Finally, we consider briefly the case that these systems are in fact binaries in which 
the sdB progenitor once filled its Roche lobe.  Several mechanisms exist 
that could work to increase the orbital eccentricity, even as mass is being lost 
or transferred, thereby preventing the binaries from circularizing.   
Most notably, these include pumping from a circumbinary disk (e.g. \citealt{der12} and references therein)
and asymmetric/phase--dependent mass loss \citep{bon08,sok00}, 
both of which have been invoked  to explain the large eccentricities observed in 
some post--AGB systems having undergone strong tidal interactions.
Long--period sdB binaries are post--RGB systems, and as such,
their sdB progenitors had significantly different luminosities, radii, envelope structures, and mass--loss rates
than AGB stars.  Eccentricity--pumping 
mechanisms operating on RGB and AGB systems will neither act on the same timescales, 
nor will they have the same ability to compete
with the effects of tides.    
We regard these scenarios, which have primarily been linked to post--AGB binaries, as uncertain
in the case of hot subdwarf binaries; further discussion 
of them is beyond the scope of this paper.
For completeness, we also recognize the action of a distant third body 
(the Kozai mechanism, \citealt{koz62}), which we regard as the
`ultimate epicycle', as another possible avenue for
increasing the eccentricity.

\section{Concluding Remarks}
\label{sec:conclusions} 
We have described an ongoing program to determine the orbital parameters
of hot subdwarfs in binaries with F--K type main sequence companions.
Preliminary results show most of our targets have orbital periods on the
order of years (rather than days), in agreement with the recent study by
\ostrv\ and the findings of \citet{gre01} from more than a decade ago.
Precise orbital solutions cannot be provided for most of our systems at
this time, owing to insufficient phase coverage.  Nonetheless, the
preliminary findings imply that the assumptions inherent in some
subdwarf formation scenarios should be revised, which probably requires
both re--tuning the parameters used in BPS models and accounting for at
least some of the long--period systems as former hierarchical triples.

%

For the three systems also observed by \ostrv, we calculated precise
orbital parameters from nearly seven years of RV measurements.  Mass
ratios were derived from the orbital RVs and are consistent with our
spectral classifications for the cool companion and canonical--mass hot
subdwarfs for two of the three systems.  In the case of the third, PG
1104+243, the derived subdwarf mass exceeds the range classically
expected for sdBs that evolved through the stable RLOF channel, implying the
subdwarf either formed from a WD merger (resulting in a
larger--than--canonical mass), or the companion has a mass different
from what we infer from its spectral type.

 A long--term goal for observers is to determine the joint distribution of 
orbital periods and eccentricities in long--period sdB binary systems.
At this time, we cannot rule out eccentric orbit solutions for PG
1104+243 or PG 1317+132.  Determining the eccentricity of PG 1317+123
will prove especially difficult due to the presence of rapid RV
variations, which might originate from star spots on the cool companion.  We
find a statistically significant eccentricity of $e=0.15 \pm 0.02$ for
PG 1338+061 and reject the circular--orbit hypothesis in this case.  The
long--period binary PG 1018$-$047 also shows evidence in favor of a
non--circular orbit \citep{dec11}, although this result needs
confirmation.  
Such departures from $e=0$, taken at face value, challenge the
prevailing notion that strong binary interactions, involving
the filling of the sdB progenitor's Roche lobe,
are always involved in the formation of hot subdwarf stars. 
If each of these binaries was (or is) a hierarchical
triple system, then the cool companion that we observe probably did not
participate directly in the formation of the hot subdwarf, in which case
its orbit would not have been circularized.
The hot subdwarf in this picture could have formed from the merger of two
He--core WDs \citep{han02,han03} or a He--core WD and a MS star \citep{cla11}.
Measurements of the rotational velocities of sdBs in long--period
systems might shed light on their formation by identifying the type
of merger.
On the other hand, if the system has always been a binary, a strong,
tidally--enhanced wind \citep{tou88} could have assisted the sdB
progenitor in ejecting its envelope on the RGB before it expanded enough
to fill its Roche lobe, thereby preserving some level of eccentricity in
the binary.  Even in the case the Roche lobe was filled previously,
 non--steady mass loss, the action of a distant third companion,
 or interactions with a circumbinary disc could have potentially 
 added eccentricity to the orbit, although these scenarios seem less likely.

In light of the updated period histogram for sdB binaries (Fig. \ref{fig:period_histogram}), 
we find it \textit{imperative} that observers continue to study both short-- and long--period
systems with all possible types of companions, to arrive at a clear understanding
of the formation channels for hot subdwarf stars.  Thus it is important to give
attention to issues of completeness and bias in our knowledge of binary orbits
(or otherwise) for sdB stars.  An important first step is to establish a more
complete catalogue or finding list, as for example in the effort described
by \citet{gir12}, who search for photometrically composite objects and use 
decomposition of the spectral energy distribution to identify binaries containing
hot subdwarfs.  We note that one of the selection criteria of \citet{gir12} is that
the binary is unresolved; this by itself does not indicate whether the binary is 
close enough to be tidally interacting, so follow--up studies are needed.
Eventually the goal is to derive full orbital elements for an unbiased sample of any
interacting binaries that are found in these surveys.  Meanwhile, work continues to
characterize the orbits of sdB binaries that are already known.  We are currently 
conducting a follow--up program with HET/HRS to determine the orbital periods, 
velocities, and eccentricities of the unsolved targets in our survey.

\begin{acknowledgements}
  This material is based upon work supported by the National Science
  Foundation under Grant No. AST--0908642. The Hobby--Eberly Telescope
  (HET) is a joint project of the University of Texas at Austin, the
  Pennsylvania State University, Stanford University,
  Ludwig--Maximilians--Universit\"{a}t M\"{u}nchen, and
  Georg--August--Universit\"{a}t G\"{o}ttingen. The HET is named in honor
  of its principal benefactors, William P. Hobby and Robert E. Eberly.
  This work is also based on observations made with the Mercator Telescope,
operated on
   the island of La Palma by the Flemish Community, at the Spanish
    Observatorio del Roque de los Muchachos of the Instituto de Astrof\'isica 
    de Canarias. 
    Finally, we thank an anonymous referee for comments and suggestions
    that helped to improve and balance discussions in the manuscript.
\end{acknowledgements}

{\it Facilities:} \facility{HET (MRS)}, \facility{Mercator1.2m (HERMES)},
\facility{KPNO:2.1m (GoldCam)}


\end{document}